\documentclass[twocolumn,aps,prb,showpacs]{revtex4}
\usepackage{epsfig}

\begin{document}


\title{Local electronic structures on the superconducting interface $LaAlO_{3}/SrTiO_{3}$}
\author{Bin Liu and Xiao Hu }

\affiliation{WPI Center for Materials Nanoarchitectonics, National
Institute for Materials Science,Tsukuba 305-0044, Japan}

\begin{abstract}

Motivated by the recent discovery of superconductivity on the
heterointerface $LaAlO_{3}/SrTiO_{3}$, we theoretically investigate
its local electronic structures near an impurity considering the influence of
Rashba-type spin-orbit interaction (RSOI) originated in the lack of
inversion symmetry. We find that local density of states near an impurity exhibits the in-gap resonance peaks due to
the quasiparticle scattering on the Fermi surface with the reversal
sign of the pairing gap caused by the mixed singlet and RSOI-induced triplet
superconducting state. We also analyze the evolutions of density of states and local
density of states with the weight of triplet pairing component determined by the strength of RSOI, which will be widely observed in thin films
of superconductors with surface or interface-induced RSOI, or various noncentrosymmetric superconductors in terms of point contact tunneling and scanning tunneling microscopy, and thus reveal an admixture of the spin singlet and RSOI-induced triplet
superconducting states.

\end{abstract}
\pacs{71.70.Ej, 73.20.At, 74.20.-z}

\maketitle


Since the discovery of a high-mobility electron gas caused by
electronic reconstruction at the interface $LaAlO_{3}/SrTiO_{3}$\cite{millis}, much attention has been paid to
its ground state. Theoretical studies suggest the charge carrier
density plays the essential role in determining the ground
state\cite{ventra}. The recent field-effect measurement on the
$LaAlO_{3}/SrTiO_{3}$ interface\cite{caviglia} indicates that
electrostatic tuning of the carrier density allows an on/off
switching of superconductivity, and drives a quantum phase
transition between a two-dimensional superconducting (SC)
state\cite{reyren} and an insulating state. This is analogous
to the case of the cuprates superconductors, where superconductivity
occurs when doping hole or electron into a Mott-insulator. However,
in contrast to chemical doping, the field-effect experiment only
modifies the charge, revealing directly the relationship between
carrier density and transition temperature $T_{c}$\cite{ahn}.
Therefore, the discovery of superconductivity controlled by electric
field is helpful for understanding the pendent mechanism of the
superconductivity, and opens the way to developing the new
mesoscopic SC circuits.

Among many interesting questions the most important one concerns the
underlying symmetry of the SC order parameter (OP). Compared with
the conventional bulk $SrTiO_{3}$ superconductor with $T_{c}\simeq
0.4 K$\cite{schooley}, the SC condensation temperature at the
interface $LaAlO_{3}/SrTiO_{3}$ is only $0.2K$\cite{reyren},
suggesting the different types of superconductivity in bulk and
interface. In particular, due to the lack of inversion symmetry
along the direction perpendicular to the interface\cite{millis},
there is a nonzero potential gradient $\nabla V$ averaged in the
unit cell, which leads to the so-called Rashba-type spin-orbit
interaction (RSOI)\cite{rashba}. The resulting RSOI changes the
nature of single-electron states, namely it leads to the lifting of
spin degeneracy and the splitting of the energy bands. In this case,
superconductivity is weakened or even suppressed according to
Anderson's theorem\cite{anderson}, which may explain the lower
$T_{c}$ in the interface compared to the bulk. Another key point is
that the RSOI induced by broken inversion symmetry breaks the
parity, the mixed singlet and triplet SC states may be possible.
This characteristic feature is the same as the non-centrosymmetric
superconductors, such as $CePt_{3}Si$\cite{bauer},
$CeRhSi_{3}$\cite{kimura},
$Li(Pd_{1-x},Pt_{x})_{3}B$\cite{togano,togano1}, where a mixing of
the spin-singlet and triplet states has been discussed due to the
absence of inversion symmetry\cite{onari,sigrist1,sigrist,bliu}.
Therefore, we expect a singlet-triplet mixing of pairing states can
be realized in the SC interface $LaAlO_{3}/SrTiO_{3}$.

In this paper, we investigate the local electronic structures near
an impurity considering the influence of RSOI in the SC interface,
which is expected to be especially important for distinguishing the
conventional superconductors from unconventional ones with the
variation of the sign of OP on the Fermi surface
(FS)\cite{balatsky}.

We start from a two-dimensional (2D) minimal tight-binding model
with the RSOI to describe the 2D electronic gas generated at the
interface $LaAlO_{3}/SrTiO_{3}$\cite{millis,sigrist,albina}, due to
the interface which is composed by $Ti$ $3d_{xy}$
electrons\cite{cezar}. It is given by
\begin{eqnarray}
H&=&\sum_{{\bf k}s}\varepsilon_{\bf k}c_{{\bf k}s}^{\dagger}c_{{\bf
k}s}+\lambda\sum_{{\bf k}ss'}{\bf g_{k}}\cdot{\bf
\sigma_{ss'}}c_{{\bf k}s}^{\dagger}c_{{\bf k}s'}, \label{eq:1}
\end{eqnarray}
where $c_{{\bf k}s}^{\dagger}$ ($c_{{\bf k}s}$) is the fermion
creation (annihilation) operator with spin $s$ and momentum ${\bf
k}$. Here,
\begin{eqnarray}
\varepsilon_{\bf k}=-2t(\cos(k_{x})+\cos(k_{y}))-\mu
\end{eqnarray}
is the tight-binding energy dispersion. The second term is the RSOI
interaction where $\lambda$ denotes the coupling constant and the
spin-orbital vector function ${\bf g_{k}}$ has the form of ${\bf
g_{k}}=(- \sin k_{y}, \sin k_{x}, 0)$\cite{rashba}. Then applying
the unitary $2\times2$ matrix
\begin{eqnarray}
U&=&\frac{1}{\sqrt{2}}\left (\matrix{1 &1\cr \frac{\bf g_{{k}1}+i\bf
g_{{k}2}}{|{\bf g_{k}}|} &-\frac{\bf g_{{k}1}+i\bf g_{{k}2}}{|{\bf
g_{k}}|}\cr}\right),
\end{eqnarray}
we can diagonalize Eq. (1) into the band representation
\begin{eqnarray}
H=\sum_{{\bf k}\nu}\xi_{{\bf k}\nu}a_{{\bf k}\nu}^{\dagger}a_{{\bf
k}\nu}
\end{eqnarray}
with the band dispersion
\begin{eqnarray}
\xi_{\bf k\pm}=\varepsilon_{\bf k}\pm\lambda|{\bf g_{k}}|.
\end{eqnarray}
As shown in Fig. 1, the RSOI term lifts the spin degeneracy by
generating two bands with reversal spin orientation.

\begin{figure}[ht]
\epsfxsize=3.5in\centerline{\epsffile{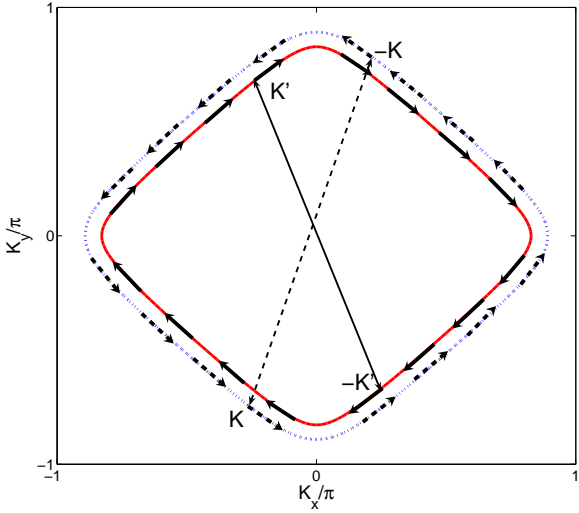}} \caption{ (color
online) Fermi surfaces for ${\bf g_{k}}=(- \sin k_{y}, \sin k_{x},
0)$ with reversal spin orientation at $\lambda/t=0.1$. The solid and
dotted double arrows denote the Cooper pairing within each Fermi
surface.}
\end{figure}

In the superconducting state, the presence of RSOI breaks the parity
and, therefore, mixes the singlet (even parity) and triplet (odd
parity) Cooper-pairing states. We parametrize the $2\times2$
pairing-potential matrix by
\begin{eqnarray}
\Delta_{\bf k}&=&(\Delta_{s}\sigma_{0}+{\bf d_{k}}\cdot{
\mathbf{\sigma}})(i\sigma_{2}),
\end{eqnarray}
where spin singlet $\Delta_{s}$ (s-wave) is assumed reasonably up to
the interface because of the conventional BCS bulk $SrTiO_{3}$
superconductor\cite{schooley}. According to weak coupling
calculations in the non-centrosymmetric
superconductors\cite{sigrist}, the RSOI-induced triplet $\bf d_{k}$
vector parallel to ${\bf g_{k}}$ gives the highest $T_{c}$ as long
as the pairing interaction stabilizes the gap function with the same
momentum dependence as that of ${\bf g_{k}}$. Thus, we define ${\bf
d_{k}}=d_{0}{\bf g_{k}}/|{\bf g_{k}}|$ in the following
calculations. Then the mean field BCS Hamiltonian has the matrix
form
\begin{eqnarray}
H_{\bf k}=\left (\matrix{\varepsilon_{\bf k} &\lambda g^{\ast}_{k}
&-d^{\ast}_{k} &\Delta_{s}\cr \lambda g_{k} &\varepsilon_{\bf k}
&-\Delta_{s} &d_{k}\cr -d_{k} &-\Delta^{\ast}_{s} &-\varepsilon_{\bf
k} &\lambda g_{k}\cr \Delta^{\ast}_{s} &d^{\ast}_{k} &\lambda
g^{\ast}_{k} &-\varepsilon_{\bf k}\cr}\right)
\end{eqnarray}
with complex notations $g_{k}=\bf g_{{k}1}+i\bf g_{{k}2}$,
$d_{k}=\bf d_{{k}1}+i\bf d_{{k}2}$, and corresponding complex
conjugates $g^{\ast}_{k}$, $d^{\ast}_{k}$. Finally the
single-particle Green's function is obtained as
\begin{eqnarray}
g({\bf k},i\omega_{n})=\left (\matrix{G({\bf k},i\omega_{n}) &F({\bf
k},i\omega_{n})\cr F^{\dagger}({\bf k},i\omega_{n}) &-G^{t}(-{\bf
k},-i\omega_{n})\cr}\right)
\end{eqnarray}
where
\begin{mathletters}
\begin{eqnarray}
G({\bf k},i\omega_{n})&=&\sum_{\tau=\pm1}\frac{1+\tau({\bf
\vec{g}_{{\bf k}}}\cdot{\bf
\sigma})}{2}G_{\tau}({\bf k},i\omega_{n}), \\
F({\bf k},i\omega_{n})&=&\sum_{\tau=\pm1}\frac{1+\tau({\bf
\vec{g}_{\bf k}}\cdot{\bf \sigma})}{2}i\sigma_{2}F_{\tau}({\bf
k},i\omega_{n}),
\end{eqnarray}
\end{mathletters}
and
\begin{mathletters}
\begin{eqnarray}
G_{\tau}({\bf k},i\omega_{n})&=&\frac{i\omega_{n}+\xi_{\bf k \tau}}{(i\omega_{n})^{2}-E^{2}_{{\bf k}\tau}}, \\
F_{\tau}({\bf
k},i\omega_{n})&=&\frac{\Delta_{\tau}}{(i\omega_{n})^{2}-E^{2}_{{\bf
k}\tau}}.
\end{eqnarray}
\end{mathletters}
Here, the SC quasiparticle excitation energy is
\begin{eqnarray}
E_{{\bf k}\tau}=\sqrt{\xi_{\bf k
\tau}^{2}+|\Delta_{\tau}|^{2}},
\end{eqnarray}
 where
$\Delta_{\tau}=\Delta_{s}+\tau|{\bf d_{k}}|$ are the SC gaps on the
energy bands and thus automatically include both inter- and
intra-band pairings in the original electron operator $c_{{\bf
k}s}$, and ${\bf \vec{g}_{\bf k}}={\bf g_{\bf k}}/{|{\bf g_{\bf
k}}|}$ is the unit vector. Then we get the density of states (DOS)
\begin{eqnarray}
\rho(\omega)=-\frac{1}{\pi}{\rm Im}\sum_{\tau, {\bf k}}G_{\tau}({\bf
k},i\omega_{n})|_{i\omega_{n}\rightarrow\omega+i0^{+}}.
\end{eqnarray}

\begin{figure}[ht]
\epsfxsize=3.5in\centerline{\epsffile{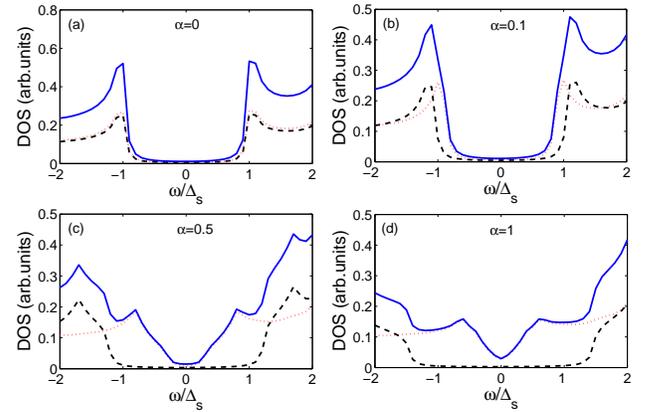}} \caption{(color
online) Evolution of DOS with the ratio $\alpha=d_{0}/\Delta_{s}$
between singlet $s$-wave and triplet pairing gaps. The dashed and
dotted curve denote the contribution of the different bands and the
solid curve refers to the total DOS.  }
\end{figure}
In Fig. 2, we show the calculated DOS for the ratio $\alpha=d_{0}/\Delta_{s}$ between the magnitudes of triplet and singlet OP ranging from 0 to 1, since the singlet pairing component is expected to be dominant near half filling compared to the RSOI-induced triplet pairing in the interface $LaAlO_{3}/SrTiO_{3}$\cite{yada}. In the absence of RSOI (Fig. 2a), namely for
zero value of the triplet pairing component, the total SC gap is
purely determined by the singlet $s$-wave gap without node on the
FS. The ``U"-shaped DOS structure is the same for both bands, and is a typical
feature of the conventional BCS superconductors. Upon introducing
and increasing the weight of the anisotropic triplet pairing component,
one finds that the total SC gap in one of the bands increases with
the value $\Delta_{s} + |{\bf d}_{\bf k}|$ while it decreases
effectively for the other band for which the total gap is
$\Delta_{s}-|{\bf d}_{\bf k}|$. Therefore the DOS in both bands are
still gapped near the Fermi energy (Fig. 2b and 2c). When the singlet $s$-wave and triplet pairing SC gaps are the same, the
accidental node forms at one of the bands and the DOS changes to a linear behavior at low energy reflecting the
formation of the line of node (Fig. 2d). We propose that point contact tunneling can probe the DOS so that reveal the mixed singlet and triplet pairing states. The corresponding momentum
dependence of the mixed singlet and triplet pairing gap functions for $\Delta_{-}$ has been plotted in Fig. 3 for $\alpha=0.5$ and $\alpha=1$ ($\Delta_{+}$ is always positive without node on
the FS and thus not plotted here). It is clearly shown that the line node has occurred at
sufficient large $\alpha=1$ (dashed line in Fig. 3b).

\begin{figure}[ht]
\epsfxsize=3.5in\centerline{\epsffile{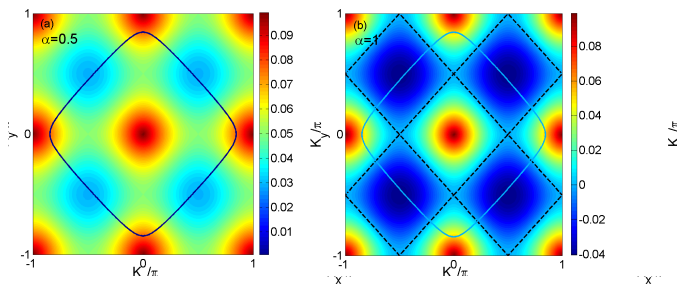}} \caption{(color
online) Mixed singlet and triplet pairing gap functions for
$\Delta_{s}-|{\bf d}_{\bf k}|$ in the first Brillouin zone at (a)
$\alpha=0.5$ and (b) $\alpha=1$. The solid line denotes Fermi
surface of band $\xi_{\bf k-}$, and the dashed line in (b) indicates
the formation of the line of node.}
\end{figure}

In order to detect the sign reversal pairing in the mixed singlet
and triplet pairing SC interface $LaAlO_{3}/SrTiO_{3}$, we calculate the local density of states (LDOS) in the presence of a
single impurity site. The impurity scattering is given by
\begin{eqnarray}
H_{\mathrm
{imp}}=V_{0}\sum_{\sigma}c_{0\sigma}^{\dagger}c_{0\sigma},
\end{eqnarray}
where without loss of generality we have taken a single-site
nonmagnetic impurity of strength $V_{0}$ located at the origin. Then
the site dependent Green's function can be written in terms of the
T-matrix formulation\cite{balatsky} as
\begin{eqnarray}
\zeta(i,j;i\omega_{n})&=&\zeta_{0}(i-j;i\omega_{n}) \nonumber\\
&+&\zeta_{0}(i,i\omega_{n})T(i\omega_{n})\zeta_{0}(j,i\omega_{n}),
\end{eqnarray}
where
\begin{eqnarray}
T(i\omega_{n})&=&\frac{V_{0}\rho_{3}}{1-V_{0}\rho_{3}\zeta_{0}(0,0;i\omega_{n})}\\
\zeta_{0}(i,j;i\omega_{n})&=&\frac{1}{N}\sum_{\bf k}e^{i\bf
k\cdot\bf R_{ij}}g(k,i\omega_{n}),
\end{eqnarray}
with $\rho_{i}$ being the Pauli spin operator, and $\bf R_{i}$ the
lattice vector, $\bf R_{ij}=\bf R_{i}-\bf R_{j}$. Finally, the LDOS
which can be measured in the scanning tunneling microscopy
experiment has been obtained as
\begin{eqnarray}
\rho(r,\omega)=-\frac{1}{\pi}\sum_{i}{\rm
Im}\zeta_{ii}(r,r;\omega+i\eta),
\end{eqnarray}
where $\eta$ denotes an infinitely small positive number.
\begin{figure}[ht]
\epsfxsize=3.5in\centerline{\epsffile{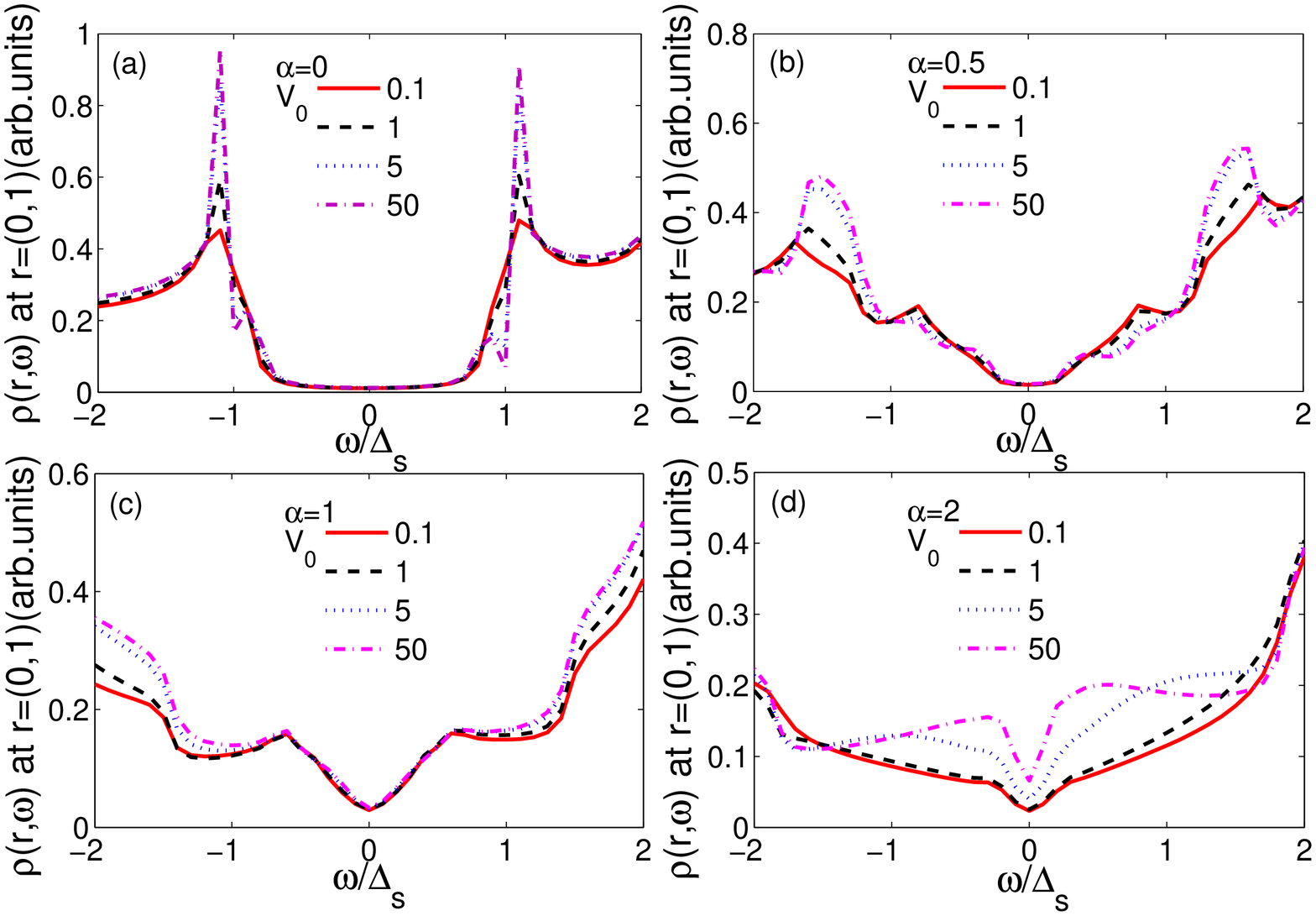}} \caption{(color
online) Evolution of LDOS near a nonmagnetic impurity with $\alpha$
for various scattering strengths $V_{0}$. }
\end{figure}

In Figs. 4a-4d we display the LDOS near a nonmagnetic impurity for
various weight ratios $\alpha$ and scattering strengths $V_{0}$.
Obviously, in the case of pure singlet s-wave pairing, the LDOS only
has two impurity resonance peaks at gap edges $\pm\Delta_{s}$ for
any scattering strength, which is known as Yu-Shiba-Rusinov
states\cite{yu}. Upon increasing the weight of triplet pairing, the
impurity resonance peaks at $\pm\Delta_{-}$ shift to low energies
when the $|\Delta_{-}|$ decreases to 0 (Fig. 4b). In Fig. 4c,
although the line node has formed on the FS, the rather small gap
value $|\Delta_{-}|=0.04\Delta_{s}$ makes the impurity-induced
resonance peaks visible only for very large scattering strength
$V_{0}$. When the triplet pairing component is dominant ($\alpha=2$
in Fig. 4d), the in-gap impurity resonance states are clearly shown.
These impurity resonance states are originated in the Andreev's
bound states\cite{balatsky} due to the quasiparticle scattering on
the FS with the reversal sign of the pairing gap. Since the triplet
paring is induced by RSOI and the magnitude is determined by the
strength of RSOI dependent of materials\cite{yada,dagan}, the above
evolution of LDOS with the weight of triplet pairing component is
expected to widely take place in thin films of superconductors with
interface or surface-induced RSOI or various noncentrosymmetric
superconductors.

\begin{figure}[ht]
\epsfxsize=3.5in\centerline{\epsffile{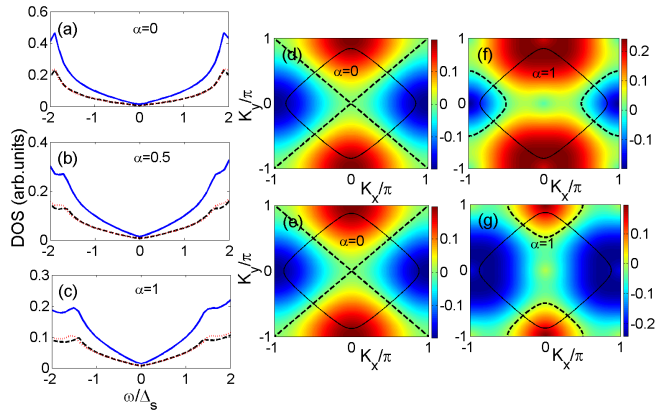}} \caption{ (color
online) Evolution of DOS for various ratio $\alpha=d_{0}/\Delta_{s}$
between singlet $d_{x^{2}-y^{2}}$-wave and triplet Cooper-pairing
states in (a), (b), and (c). The dashed and dotted curve denote the
contribution of the different bands and the solid curve refers to
the total DOS. The corresponding singlet and triplet pairing gap
functions in the first Brillouin zone at $\alpha=0$ for $\Delta_{+}$
(d) and $\Delta_{-}$ (e), and $\alpha=1$ for $\Delta_{+}$ (f) and
$\Delta_{-}$ (g), where the solid line denotes corresponding Fermi
surface of band $\xi_{\bf k+}$ in (d) and (f), and band $\xi_{\bf
k-}$ in (e) and (g), and the dashed line indicates the node line.}
\end{figure}
\begin{figure}[ht]
\epsfxsize=3.5in\centerline{\epsffile{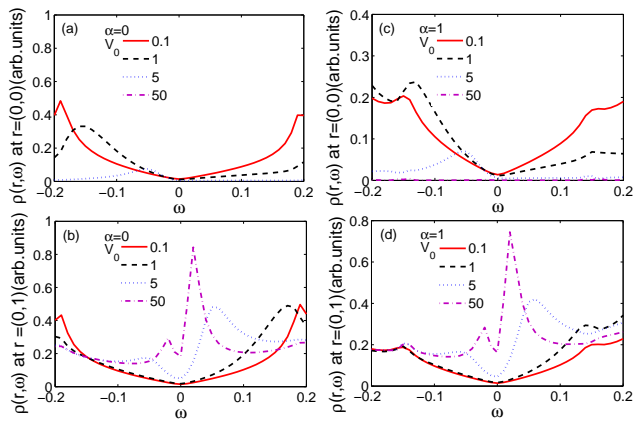}} \caption{(color
online) LDOS near an impurity for $\alpha=0$ and $\alpha=1$ with
various scattering strengths $V_{0}$.}
\end{figure}
We also investigate the possible mixed singlet d-wave and triplet
pairing states in the superconducting interface
$LaAlO_{3}/SrTiO_{3}$\cite{yada}. In Fig. 5, we plot the DOS and
corresponding momentum-dependent gap functions in the mixed singlet
$d_{x^{2}-y^{2}}$-wave and triplet pairing states. It is shown that
upon increasing the weight of triplet pairing, the DOS always keeps
a ``V"-shape behavior at the low energies (Figs. 5a-5c) due to the
existence of line node on the FS seen in Figs. 5d-5g. Although
RSOI-induced triplet pairing gap changes the shape of pure
$d_{x^{2}-y^{2}}$-wave gap function, the nodal line on the FS always
persists. The LDOS near an impurity in mixed singlet d-wave and
triplet pairing states has been plotted in Fig. 6. We find in-gap
impurity resonance states for different values of $\alpha$, similar
to that of the zero bias resonance peak on the Zn impurity in
cuprates superconductors\cite{pan}. Compared to the case of
coexisting singlet $s$-wave and triplet pairing states, the
evolutions of DSO and LDOS in the mixed singlet d-wave and triplet
pairing states with the weight of triplet pairing component exhibits
different RSOI's influence on the electronic structures near an
impurity, and thus can be easily differentiated by point contact
tunneling or scanning tunneling microscopy.


In summary, we study the mixed singlet and triplet cooper pairing
states on the interface $LaAlO_{3}/SrTiO_{3}$ based on a minimal tight-binding model considering the influence of RSOI
induced by the lack of inversion symmetry. Applying T-matrix
approximation, we theoretically investigate its impurity induced
resonance states. We find that local density of states near an
impurity exhibits the in-gap resonance peaks due to
the quasiparticle scattering on the FS with the reversal sign of the
pairing gap caused by the mixed singlet and RSOI-induced
triplet cooper pairing SC state. We also reveal the evolutions of
DOS and LDOS with the weight of triplet pairing component. These features will be widely observed via point contact tunneling and scanning
tunneling microscopy in thin films of
superconductors with interface or surface-induced RSOI or various
superconductors without inversion symmetry.

Recently, the observation of superconductivity in a topological
insulator $Bi_{2}Se_{3}$ has attracted much interest on its
topological surface states due to prominent role played by
spin-orbit interaction\cite{hor}. Its unconventional
superconductivity has been discussed in a recent paper\cite{linder}
where possible singlet or triplet pairing states was proposed. Our
present approach can be directly applied to investigate its pairing
symmetry and superconductivity.

Recently, we notice the magnetotransport experiment
reports\cite{caviglia1} that the RSOI is caused by the lack of
inversion symmetry on the interface $LaAlO_{3}/SrTiO_{3}$ and its
strength can be tuned applying an external electric field.

This work was supported by WPI Initiative on Materials
Nanoarchitronics, MEXT, Japan.

\end{document}